\documentclass[twocolumn,english,amsmath,amssymb,superscriptaddress,nofootinbib]{revtex4-1}
\usepackage{mathptmx}
\usepackage[T1]{fontenc}
\usepackage[latin9]{inputenc}
\usepackage{amsmath}
\usepackage{amssymb}
\usepackage{graphicx}
\usepackage{esint}

\makeatletter
\usepackage{babel}

\usepackage{xr}
\makeatother

\usepackage{babel}
\begin{document}

\title{Amplified emission and lasing in a plasmonic nano-laser with many three-level molecules}

\author{Yuan Zhang}
\email{yzhang@phys.au.dk}

\author{Klaus M\o lmer}
\email{moelmer@phys.au.dk}

\address{Department of Physics and Astronomy, Aarhus University, Ny Munkegade
120, DK-8000 Aarhus C, Denmark}
\begin{abstract}
Steady-state plasmonic lasing is studied theoretically for a system consisting of many dye molecules arranged regularly around a gold nano-sphere. A three-level model with realistic molecular dissipation is employed to analyze the performance as function of the pump field amplitude and number of molecules. Few molecules and moderate pumping produce a single narrow emission peak because the excited molecules transfer energy to a single dipole plasmon mode by amplified spontaneous emission. Under strong pumping, the single peak splits into broader and weaker emission peaks because two molecular excited levels interfere with each other through coherent coupling with the pump field and with the dipole plasmon field. A large number of molecules gives rise to a Poisson-like distribution of plasmon number states with a large mean number characteristic of lasing action. These characteristics of lasing, however, deteriorate under strong pumping because of the molecular interference effect.
\end{abstract}

\maketitle

\section{Introduction}

The prospects to observe quantum effects in the response of metallic material to light \cite{MSTame}
and explore their potential applications \cite{RMMa-0,MPelton,PBerini} have stimulated abundant recent studies in plasmonics. Most of the envisioned applications are based on the enhanced and localized electromagnetic
(EM) field near metallic films and nano-particles (MNP),
due to the formation of surface plasmons, \textit{i.e.}, collective oscillations
of the EM field and the conduction electrons in the metal. This enhancement
leads to strong light-matter interaction and enhanced absorption \cite{NICade,YZelinskyy}, emission \cite{PAnger}, and Raman
scattering \cite{MFleischmann,PJohansson} of quantum emitters, which 
can be naturally utilized to improve sensitivity of spectroscopic
instruments \cite{SYDing, SNie} and efficiency of light emitting diodes \cite{XFGu,NGao}
and solar cells \cite{HAAtwater,LJWu}. Plasmon field modes, confined within sub-wavelength volumes,  were utilized by Bergman and
Stockamn \cite{Bergman} to propose the spaser or plasmonic nano-laser  in 2003, which was then verified in experiments by  Noginov and his co-workers in 2009 \cite{MANoginov}
with a system consisting of a gold nano-sphere and many dye molecules.
Later experimental demonstrations of the spaser have been reported with structures like semiconductor wires \cite{JHo,YHChou,BTChou} and squares \cite{RMMa,RMMa-1} on metallic films, with semiconductor pillars \cite{MKha,KDing}, dots
\cite{AMatsudira,CYLu}, and wires \cite{CYLu-1,SWChang} inside metallic
shells and with dye molecules deposited in periodically arranged MNP arrays
\cite{JYSuh,WZhou,AYang,AYang-1,AHSchokker}.

\begin{figure}
\begin{centering}
\includegraphics[scale=1.0]{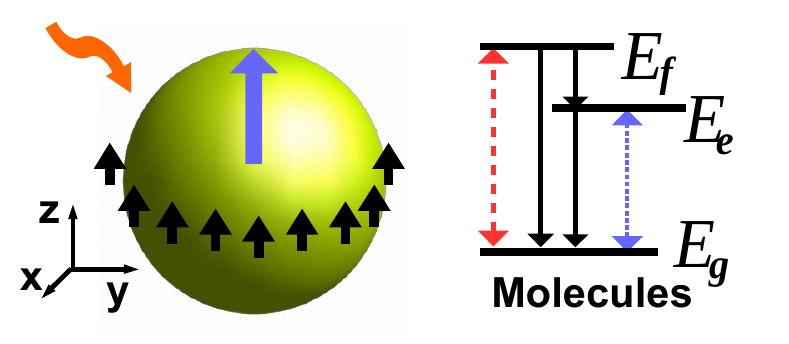}
\par\end{centering}
\caption{\label{fig:scheme-three-level} The figure depicts a
gold nano-sphere ($10$ nm radius) is surrounded by dye molecules along
the equator about $2.5$ nm from the sphere-surface. The molecules
have transition dipole moments along the $z$-axis (black arrows) and
couple only with the dipole plasmon $z-$mode (blue arrow). The molecules are assumed identical with three energy levels ($E_{g}$, $E_{e}$, $E_{f}$), and are coupled coherently to an external optical pump field (red dashed arrow) and to the dipole plasmon (blue dotted arrows), and decay by dissipation (black arrows). }
\end{figure}

Semi-classical theories based on rate equations \cite{Bergman,AMatsudira,CYLu,CYLu-1} and  Maxwell-Bloch equations \cite{WZhou,AYang,AYang-1} have been utilized to study the plasmonic nano-laser. These equations can incorporate coherent molecular excitation mechanisms, inhomogeneous coupling and multiple lasing modes but they ignore quantum correlations between the molecules and the lasing modes. While quantum theories accounting for these correlations have been presented with plasmon number states \cite{MRichter} or coherent states \cite{VMPar}, the molecules have been always considered as identical two-level systems and the pumping mechanism has been simply modeled with an incoherent pumping rate.

To model the real and coherent pumping mechanism more realistically and understand its influence on  the system quantum properties, we develop a quantum laser theory for the system, shown in Fig. \ref{fig:scheme-three-level}(a) (resembling the one in the experiment \cite{MANoginov}). We thus consider the molecules as identical three-level systems,  where one molecular transition couples with an external optical (driving) field and another transition couples with the plasmon lasing mode. Remarkably, we find that the two molecular excited states can interfere with each other due to the two couplings, which leads to emission peak splitting and reduced emission intensity. To our knowledge, this effect has not been predicted by previous theories. 

The article is outlined as follows. In Sec. \ref{sec:master-equation}, we introduce a master equation to describe the dynamics of many molecules and a single quantized plasmon mode. In Sec. \ref{sec:collective-method}, we solve this equation with an exact numerical method based on a collective reduced density matrix (RDM), which applies for systems with identical molecules. In Sec. \ref{sec:approximate-photon-RDM}, we apply an approximate method, more general than  Lamb's laser theory \cite{MSargent}, based on elimination of the molecular degrees of freedom and solution of a plasmon master equation. This method is verified and subsequently used to simulate systems with hundreds of molecules. In Sec.\ref{sec:conclusions}, we conclude and discuss possible extensions of our theory.

\section{Master Equation \label{sec:master-equation}}

We assume the molecules located around the equator of the sphere at a distance of $2.5$ nm from the surface, see Fig. \ref{fig:scheme-three-level}, which is sufficient to exclude tunneling ionization  \cite{KJSavage}. The molecules thus only exchange energy with the sphere through the dipole
plasmon mode, which is resonant with the molecules and has its transition
dipole moment along the $z$-axis. The more general configuration involving randomly oriented molecules and three resonant dipole plasmons has been investigated in \cite{YZhang-4}.
Higher multipole plasmons are off-resonant and do not affect the molecules \cite{YZhang-3} apart from a contribution to their excited state decay \cite{JGersten}.

The master equation for the density operator $\hat{\rho}$ of the plasmon mode and the molecular emitters reads:
\begin{equation}
\frac{\partial}{\partial t}\hat{\rho}=-\frac{i}{\hbar}\left[H_{{\rm pl}}+H_{{\rm m}}+V_{{\rm pl-m}}+V_{{\rm m}}\left(t\right),\hat{\rho}\right]-\mathcal{D}\left[\hat{\rho}\right].\label{eq:RDO}
\end{equation}
The plasmon mode with excitation energy $\hbar\omega_{\mathrm{pl}}$ is
described by  the Hamiltonian $H_{{\rm pl}}=\hbar\omega_{\mathrm{pl}}C^{+}C$ with bosonic creation and annihilation operators $C^{+}$, $C$. The Hamiltonian of $N_{\rm m}$ molecules reads $H_{{\rm m}}=\sum_{n=1}^{N_{{\rm m}}}\sum_{a_{n}} E_{a_n} \left|a_{n}\right\rangle \left\langle a_{n}\right|$, with the single molecular
ground state $\left|a_{n}=g_{n}\right\rangle $,
and first $\left|a_{n}=e_{n}\right\rangle $ and second $\left|a_{n}=f_{n}\right\rangle $
excited states with energies $E_{g_n}$,$E_{e_n}$,$E_{f_n}$, respectively.
We assume that the plasmon mode couples resonantly with the molecular ground-to-first
excited state transition through the interaction Hamiltonian $V_{{\rm pl-m}}=\hbar\sum_{n=1}^{N_{{\rm m}}}v_{ge}^{\left(n\right)}\left(C^{+}\left|g_{n}\right\rangle \left\langle e_{n}\right|+{\rm h.c.}\right)$
(in the rotating wave approximation). Here, the coefficient $\hbar v_{ge}^{\left(n\right)}=[\mathbf{d}_{ge}^{\left(n\right)}\cdot\mathbf{d}_{\mathrm{pl}}-3 (\mathbf{d}_{ge}^{\left(n\right)}\cdot\hat{\mathbf{x}}_{n})\left(\mathbf{d}_{\mathrm{pl}}\cdot\hat{\mathbf{x}}_{n}\right)]/\left|\mathbf{X}_{n}\right|^{3}$
is determined by molecular and plasmon transition dipole moments $\mathbf{d}_{ge}^{\left(n\right)}$,
and $\mathbf{d}_{\mathrm{pl}}=d_{\mathrm{pl}}\mathbf{e}_{z}$
($\mathbf{e}_{z}$ is the unit vector along the $z$-axis), and by the 
vectors $\mathbf{X}_{n}=X_{n}\hat{\mathbf{x}}_{n}$, connecting the
molecules and the sphere-center. We assume that the molecules are subject to a driving field with frequency
$\omega_{0}$, resonant with the molecular ground-to-second excited state transition through the Hamiltonian
$V_{\mathrm{m}}\left(t\right)=\hbar\sum_{n=1}^{N_{{\rm m}}}v_{gf}^{\left(n\right)}\left(e^{i\omega_{0}t}\left|g_{n}\right\rangle \left\langle f_{n}\right|+{\rm h.c.}\right)$, where $\hbar v_{gf}^{\left(n\right)}=\mathbf{d}_{gf}^{\left(n\right)}\cdot\mathbf{n}E_{0}$
is determined by the molecular transition dipole moment $\mathbf{d}_{gf}^{\left(n\right)}$
and the driving field amplitude $E_{0}$ and polarization $\mathbf{n}$. Dissipation
is accounted for by several \emph{Lindblad} terms:
\begin{equation}
\mathcal{D}\left[\hat{\rho}\right]=\left(1/2\right)\sum_{u}k_{u}\left(\left[\hat{L}_{u}^{+}\hat{L}_{u},\hat{\rho}\right]_{+}-2\hat{L}_{u}\hat{\rho}\hat{L}_{u}^{+}\right).\label{eq:Dissipation}
\end{equation}
Plasmon damping is included by terms with $k_{u}=\gamma_{\mathrm{pl}}$
and $\hat{L}_{u}=C$. The decay processes in the individual molecules are represented
by terms with $k_{u}=k_{a\to b}^{\left(n\right)}$ and $\hat{L}_{u}=\left|b_{n}\right\rangle \left\langle a_{n}\right|$
for $E_{a_n}>E_{b_n}$. For simplicity, we ignore pure dephasing of the molecules.

\section{Collective Reduced Density Matrix Equation \label{sec:collective-method}}

Due to symmetry, the density matrix must at all times be invariant under permutation of the identical molecules. For two-level systems, this is utilized in collective representations  of the density matrix with \emph{Dicke} states \cite{UMartini,BAChase}, SU(4) group theory \cite{MXu} and collective numbers  \cite{MRichter,YZhang0}. Here, we choose the latter representation since it can be easily extended to systems with multi-level molecules \cite{MGegg,YZhang1}. We consider the density matrix element   $\rho_{\beta\nu,\alpha\mu}\left(t\right)$ between plasmon number states $\left|\mu\right\rangle$ and $\left| \nu\right\rangle$ and molecular product states $\left|\alpha\right\rangle \equiv\prod_{n=1}^{N_{{\rm m}}}\left|a_{n}\right\rangle ,\left|\beta\right\rangle \equiv\prod_{n=1}^{N_{{\rm m}}}\left|b_{n}\right\rangle $. The invariance under permutation of the molecules permits the representation of many of these matrix elements by one and the same number that we denote by  $\rho_{\bf n}^{\mu \nu}$  where ${\bf n}=\{n_{ab}\}$ counts the number of occurrences of $\left|a_n=a\right\rangle$ and $\left|b_n =b\right\rangle$ in the states $\left|\alpha\right\rangle$ and $\left|\beta \right\rangle $. The master equation couples different matrix elements which can be all systematically represented in the reduced form. We effectively obtain a significantly reduced set of coupled equations for the collective RDM $\rho_{\bf n}^{\mu\nu}$. The resulting equations are presented as Eq. \eqref{eq:symmetry-rdm} in  Appendix \ref{sec:population-emission}. The number $C_{{\rm N_m} +8}^{8}(N_{\rm pl} +1)^2$ of $\rho_{\bf n}^{\mu\nu}$  is order of magnitudes smaller than the number $(3^{N_m}\cdot(N_{\rm pl}+1))^2$ of  $\rho_{\beta\nu,\alpha\mu}$. Here, $N_{\rm pl}$ indicates the highest plasmon state considered.  For example, for a system with ten molecules and six plasmon states, we reduce the number of elements from  $1.3\times 10^{11}$  to $1.6\times 10^6$  . For more details of the method please refer to our article \cite{YZhang2} about the collective density matrix of multi-level emitters. 

Using the density matrix with elements $\rho_{\bf n}^{\mu\nu} \equiv\rho_{\beta\nu,\alpha\mu}$,  we can calculate all physical observables, for example, the population of the molecular states $\left|a = a_n \right\rangle $: $P_a$ ($a$ for $g,e,f$), the plasmon number state distribution $\left|\mu\right\rangle $: $P_{\mu}=\sum_{\alpha}P_{\alpha\mu}$, and the mean plasmon number $A_{\mathrm{pl}}=\sum_{\mu}\mu P_{\mu}$
as well as its normalized second factorial moment $g_{\mathrm{pl}}^{2}\left(0\right)=\sum_{\mu}\mu\left(\mu-1\right)P_{\mu}/A_{\mathrm{pl}}^{2}$ (this number characterizes the number distribution and equals unity for a coherent state).
We can also apply the quantum regression
theorem \cite{PMeystre} and use the master equation to calculate two-time correlation functions and, e.g., the emission spectrum. The explicit expressions to compute these observables  are detailed in  Appendix \ref{sec:population-emission}.

\subsection{Exact Numerical Results for Systems with up to Ten Molecules}

\begin{figure}[tp]
\begin{centering}
\includegraphics[scale=1.0]{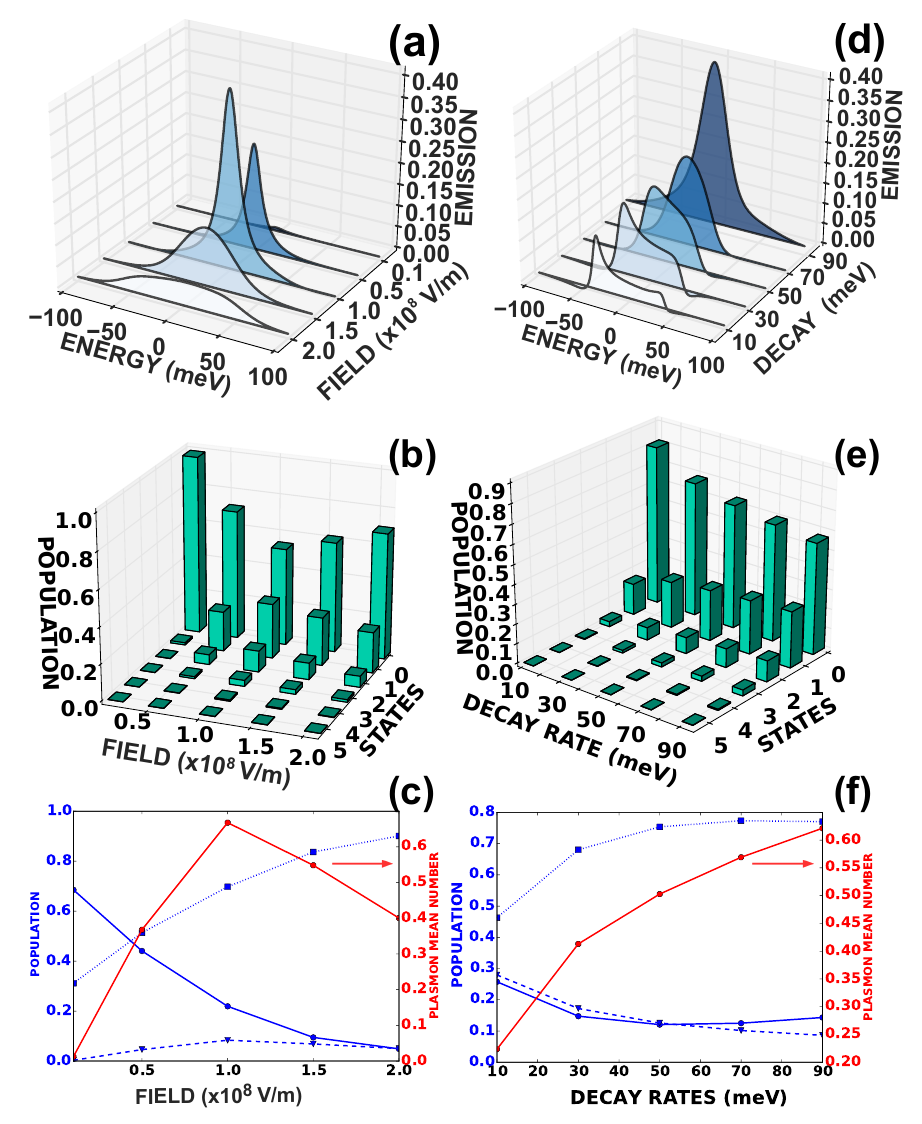}
\par\end{centering}
\caption{\label{fig:strength-decay} Steady-state properties of systems with eight molecules  for different strength of the driving field $E_0$ (panels a,b,c) and different decay rates $k_{f \to e}$ (panels d,e,f). Panels (a,d) for emission spectrum  (the photon energy is given relative to $\hbar\omega_{\mathrm{pl}}$). Panels (b,e) for plasmon state population $P_\mu$. Panels (c,f) for molecular state population $P_a$ (blue lines; solid line $P_g$, dotted line $P_e$, dashed line $P_f$) and plasmon mean number $A_{\rm pl}$. 
 All further parameters are specified in Table \ref{tab:parameters} in  Appendix \ref{sec:parameters}.}
\end{figure}

Fig.\ref{fig:strength-decay} shows the influence of strength of the driving field $E_{0}$ (panels a,b,c) and decay rate $k_{f \to e}$ (panels d,e,f) on the steady-state properties of systems with eight molecules. In the panel (a)  the emission spectrum is weak for a weak driving field and it develops a sharp peak for moderate driving due to amplified spontaneous emission, which is verified by
the increased population of the plasmon excited states (cf. panel b) and the molecular excited state (cf. the blue dotted line in panel c). At strong driving, the spectrum becomes broadened  and weak as a consequence of quantum interference between
the two molecular excited states. Although these states do not directly
couple with each other, they are coupled through the coherent
coupling with the driving field and with the plasmon mode, and a similar
phenomenon occurs in the context of lasing without inversion
\cite{JMompart}. If the fields couple to two separate
transitions as in the four-level molecular model studied
in \cite{WZhou}, this interference effect is absent, and the emission
intensity saturates but it does not deteriorate for large $E_{0}$. Similar behavior of the mean plasmon number  $A_{\mathrm{pl}}$ is observed (cf. the red line in panel c).

In Fig.\ref{fig:strength-decay}(d), the emission spectrum shows two peaks, a broad peak and a single sharp peak with a broad background, respectively, for small, moderate, and large decay rate. The two peaks arise due to the quantum interference effect, which is suppressed in the case of strong dissipation where the sharp peak results from the amplified spontaneous emission.
The emission intensity increases with increasing $k_{f\to e}$
because the molecules with increasing population on the excited state $P_{e}$ (cf. the dotted line in  panel f) transfer more energy to the plasmon, which leads to increased population of the plasmon higher excited states (cf. panel e) and plasmon mean number (cf. the red line in panel f).

\begin{figure}[tp]
\begin{centering}
\includegraphics[scale=1.0]{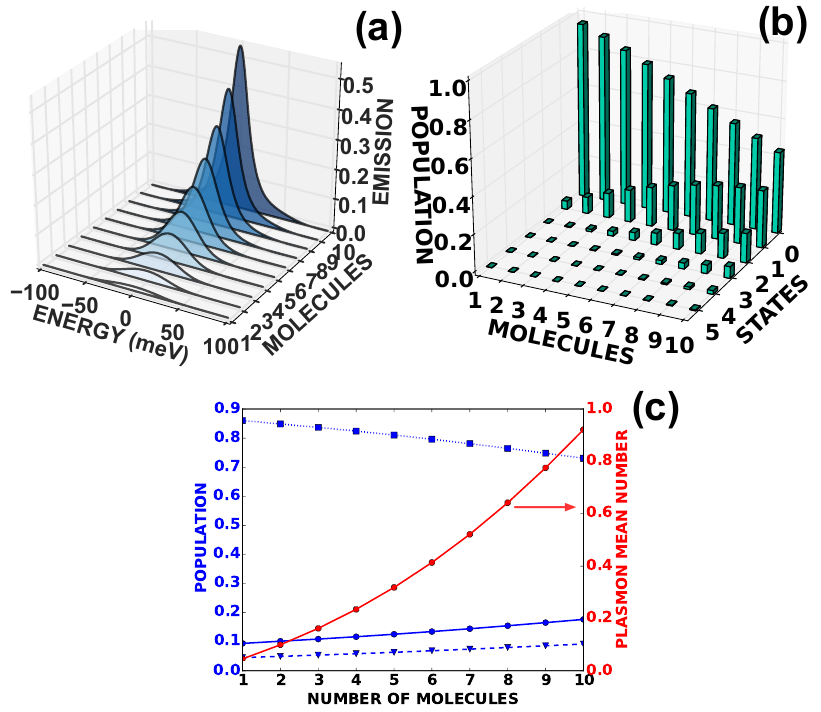}
\par\end{centering}
\caption{\label{fig:number-emitters}  Steady-state emission spectrum (panel a, the photon energy is given relative to $\hbar\omega_{\mathrm{pl}}$),  plasmon number state distribution $P_{\mu}$ (panel b) and molecular state population $P_a$ (panel c, blue lines, solid line $P_g$, dotted line $P_e$, dashed line $P_f$) as well as plasmon mean number $A_{\rm pl}$ (panel c, red line) 
for different numbers of molecules. All further parameters are specified in Table \ref{tab:parameters} in  Appendix \ref{sec:parameters}.}
\end{figure}

Fig.\ref{fig:number-emitters} displays how the number of molecules
$N_{\mathrm{m}}$ affects the emission and the plasmon state population $P_{\mu}$, respectively.
In the panel (a) the emission shows higher intensity and spectral narrowing for the larger values of $N_{\mathrm{m}}$. Panel (b) shows that only the $\mu=0,1$ plasmon number states are populated for $N_{\textrm{m}}=1$. With increasing $N_{\mathrm{m}}$, the plasmon states with $\mu\geq1$ are gradually populated indicating an increased plasmon mean number (cf. the red line in panel c) and strong plasmon excitation. However, because of the backaction of the strong plasmon excitation, the population of molecular excited state $P_e$ reduces (cf. the dotted blue line in panel c). These results clearly indicate a transition from fluorescence to amplified spontaneous emission. In the following section we shall demonstrate further increase of the plasmon excitation, i.e. lasing action, for systems with more molecules.

\section{Approximate Plasmon Reduced Density Matrix Equation \label{sec:approximate-photon-RDM}}

\begin{figure*}[tp]
\begin{centering}
\includegraphics[scale=1.1]{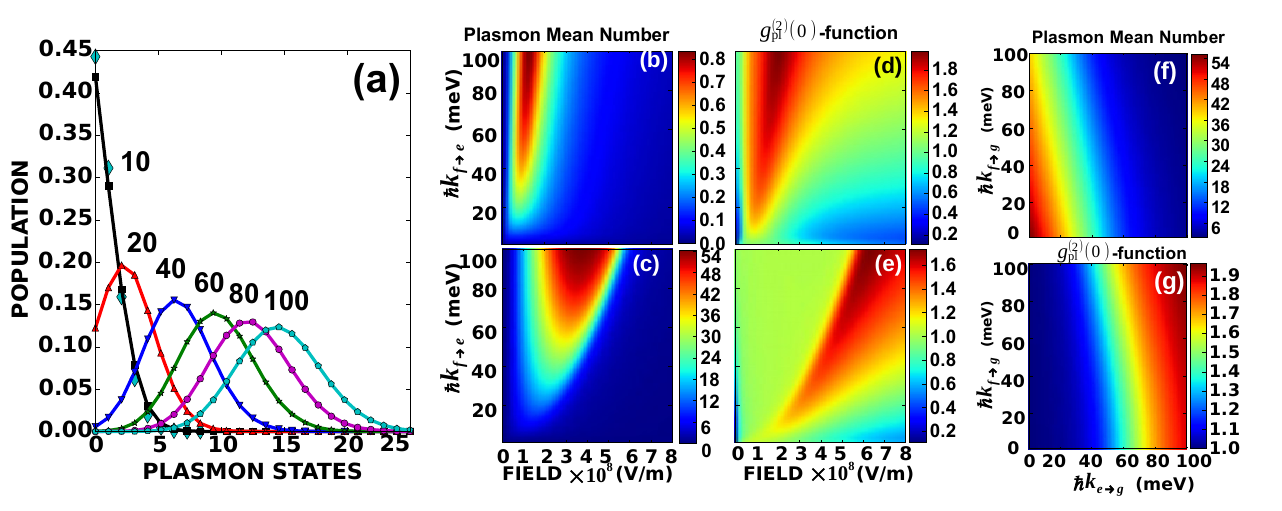}
\par\end{centering}
\caption{\label{fig:approximate-rdm} Steady-state properties of systems:   panel (a) calculated by Eq. (\ref{eq:plasmon-state-population}) shows $P_\mu$ for different number of molecules, large diamonds show the exact result for ten molecules;  panels (b,c) show $A_{\mathrm{pl}}$ and panels (d,e) show  $g_{\mathrm{pl}}^{\left(2\right)}\left(0\right)$ versus the decay rate $k_{f\to e}$ and the strength of the driving field $E_{0}$; panels (b,d) are for systems with 10 molecules; panels (c,e) are for systems with 200 molecules; panel (f) shows $A_{\mathrm{pl}}$ and (g) shows $g_{\mathrm{pl}}^{\left(2\right)}\left(0\right)$ versus decay rates $k_{e\to g}$ and $k_{f\to g}$ for systems with 200 molecules  ($\hbar k_{f\to e}=100$ meV and $E_{0}=3\times10^{8}$
V/m). Other parameters are according to Table \ref{tab:parameters}
in Appendix \ref{sec:parameters}}
\end{figure*}

The collective RDM allows only simulation of systems with up to ten molecules because of the huge number of matrix elements and thus it is necessary to develop approximate methods to
solve Eq. \eqref{eq:RDO} for larger systems. Here, we follow the same methods as have been applied to the laser \cite{MSargent,MOScully} to adiabatically eliminate the molecular
degree of freedom and derive equations
only for the plasmon RDM  $\rho_{\mu\nu}\equiv\text{tr}_{\text{S}}\left\{ \hat{\rho}\left(t\right)\left|\nu\right\rangle \left\langle \mu\right|\right\} $. The detailed derivation is given in Appendix \ref{sec:approximate-prdm}. Here, we only outline the main procedure and present the
final results.
 
Due to the molecule-plasmon coupling, the elements $\rho_{\mu\nu}$ of the reduced density matrix for the plasmon mode depend on the values of the molecule-plasmon correlations, $\rho_{e\mu-1,g\nu}^{\left(n\right)}$
and $\rho_{g\mu+1,e\nu}^{\left(n\right)}$, which are, in turn, coupled to the correlations $\tilde{\rho}_{f\mu,g\nu}^{\left(n\right)}\equiv e^{i\omega_{0}t}\rho_{f\mu,g\nu}^{\left(n\right)}$
and $\tilde{\rho}_{f\mu,e\nu-1}^{\left(n\right)}\equiv e^{i\omega_{0}t}\rho_{f\mu,e\nu-1}^{\left(n\right)}$, 
due to the simultaneous coupling with the
plasmon and driving fields.
Since the dissipation of both the molecules and the plasmon contributes to the decay of
$\rho_{a\mu,b\nu}^{\left(n\right)}$, they should reach steady-state much faster than the reduced elements $\rho_{\mu\nu}$, that evolve mainly by the plasmon
decay rate. Thus, we apply an adiabatic elimination by assuming the steady state solution for $\rho_{a\mu,b\nu}^{\left(n\right)}$. Similarly, correlations $\rho_{e\mu,e\nu}^{\left(n\right)}$,
$\rho_{g\mu-1,g\nu-1}^{\left(n\right)}$ and $\rho_{f\mu-1,f\nu-1}^{\left(n\right)}$ can
be expressed as combinations of $\rho_{e\mu-1,g\nu}^{\left(n\right)}$
and $\rho_{g\mu+1,e\nu}^{\left(n\right)}$ and $\rho_{\mu\nu}$, leading eventually to a closed set of equations for the plasmon reduced density matrix, cf. Eq. \eqref{eq:plasmonRDM-final}
in Appendix \ref{sec:approximate-prdm}. It turns out that we obtain separate equations for the diagonal and off-diagonal elements, and we focus here on the plasmon state populations $P_{\mu}=\rho_{\mu\mu}$, obeying
\begin{eqnarray}
\frac{\partial}{\partial t}P_{\mu} & = & -\left[\left(\gamma_{\text{pl}}\mu+k_{\mu}\right)P_{\mu}-p_{\mu}P_{\mu-1}\right]\nonumber \\
 & + & \left(\gamma_{\text{pl}}\left(\mu+1\right)+k_{\mu+1}\right)P_{\mu+1}-p_{\mu+1}P_{\mu}.\label{eq:plasmon-state-population}
\end{eqnarray}
The effective rates $k_{\mu}$ and $p_{\mu}$ are defined by Eqs. \eqref{eq:moleculePD}
and \eqref{eq:moleculePP} in Appendix \ref{sec:approximate-prdm} and they can be viewed as extended Einstein $A$ and $B$ coefficients due to the molecular pumping mechanism. The rate $k_{\mu}$ describes loss of plasmons towards molecular excitation, and causes depopulation of higher excited plasmon states and increased population of lower excited plasmon states. The rate $p_{\mu}$ describes the plasmon emission by the excited molecules.

In steady-state the time-derivatives in Eq. \eqref{eq:plasmon-state-population} vanish, and we obtain a recursion relation for the populations
\begin{equation} \label{eq:recursive}
r_{\mu}=P_{\mu}/P_{\mu-1}=p_{\mu}/\left(\gamma_{\text{c}}\mu+k_{\mu}\right).
\end{equation}
Together with the normalization $\sum_{\mu}P_{\mu}=1$, this relation allows us to readily calculate $P_{\mu}$ for system with hundreds and even thousands of molecules.

\subsection{Results for Systems with Hundreds of Molecules }
The recursion relation (\ref{eq:recursive}) reproduces all the exact results shown in Fig.\ref{fig:number-emitters}(b) very well and thus proves the validity of the adiabatic elimination of the molecular degrees of freedom in large systems. In this section we thus apply the reduced master equation \eqref{eq:plasmon-state-population} and thus Eq.(\ref{eq:recursive}) to systems with hundreds of molecules, cf. Fig.\ref{fig:approximate-rdm}. The panel (a) shows the plasmon number distribution $P_{\mu}$ for systems with different numbers $N_{\textrm{m}}$ of molecules. The big diamonds are the exact result for ten molecules and agree very well with the black squares calculated with Eq.(\ref{eq:recursive}). For $N_{\textrm{m}}=10$ and $20$, $P_{\mu}$ decreases with increasing $\mu$, while for $N_{\textrm{m}}\geq40$, the population
distributions show a peak-structure, shifting
to higher plasmon excitation with increasing $N_{\mathrm{m}}$. The Poisson-like distributions
indicate lasing action, associated with the formation of a coherent state.

The variation of $A_{\mathrm{pl}}$
and $g_{\mathrm{pl}}^{\left(2\right)}\left(0\right)$ is shown as function of the driving field $E_{0}$ and the molecular dissipation rate $k_{f\to e}$  for systems with $N_{\rm m}=10$  molecules in the upper panels (b,d) and $N_{\rm m} = 200$ in the lower panels (c,e) in Fig.\ref{fig:approximate-rdm}. Note the different plasmon number color bars in the panels (b,c). The smaller
systems show amplified spontaneous emission and their statistics is close to thermal,  $g_{\mathrm{pl}}^{\left(2\right)}\left(0\right) \approx 2$,
,cf. panel (d), when the plasmons are excited. The larger systems show lasing action with a large mean plasmon number $A_{\mathrm{pl}}\gg1$, and a Poisson-like distribution with  $g_{\mathrm{pl}}^{\left(2\right)}\left(0\right)=1$,
cf. panel (e).

Fig.\ref{fig:approximate-rdm} (f,g) illustrate the influence of
the decay rates $k_{f\to g},k_{e\to g}$ for systems with $200$ molecules
($E_{0}=3\times10^{8}$ V/m and $\hbar k_{f\to e}^{\left(n\right)}=100$
meV). The panel (f) shows that the increased decay rates reduce $A_{\mathrm{pl}}$. This occurs because 
they reduce the population inversion $P_{e}-P_{g}$ and because $k_{e\to g}$ contributes to the molecular dephasing rate $\gamma_{ge}^{\left(n\right)}$. The
panel (g) shows that regimes of high (low) plasmon numbers are 
governed by Poissonian (super Poissonian) statistics. 

\section{Discussion and Outlook\label{sec:conclusions}}

In summary, we have solved the master equation for a plasmonic nano-laser with three-level molecules under continuous optical pumping. We developed and applied an exact method for small systems with up to ten molecules and an approximate method for larger systems with hundreds of molecules. The small systems show amplified spontaneous emission, indicated by
an increased emission intensity, but predominant population of lower
plasmon number states and thermal like statistics. The systems show a destructive
quantum interference effect for strong pumping, leading to a reduced emission intensity
and a split emission spectrum. The larger systems  show lasing action as witnessed by a population of higher plasmon number states and Poisson-like statistics. The exact method can be generalized to systems with identical multi-level emitters \cite{YZhang2} and is thus ideal to explore collective effects in those systems, for example, collective strong coupling and superradiance. 

\section*{ACKNOWLEDGMENTS}
Y. Z. and K. M. acknowledge Chuan Yu and Lukas F. Buchmann for
several illuminating discussions. This work was supported by Villum
Foundation (Y. Z. and K. M.).

%%%%%%%%%%%%%%%%%%%%%%%%%%%%%%%%%%%
%% 								appendix 
%%%%%%%%%%%%%%%%%%%%%%%%%%%%%%%%%%%
\appendix

\section{System Parameters \label{sec:parameters}}

In Table \ref{tab:parameters}, we collect the reference parameters
for our simulations. We consider a gold nano-sphere of  $10$ nm radius. For such a nano-sphere, there are three dipole plasmons with transition dipole moments along three axes of Cartesian
coordinate system. They have the same excitation energy $\hbar\omega_{\mathrm{pl}}=2.6$ eV and damping rate $\hbar\gamma_{\mathrm{pl}}=100$ meV as well as a transition dipole
moment $d_{\mathrm{pl}}=2925$ D. The driving field has an energy
of $\hbar\omega_{\mathrm{c}}=2.7$ eV and varying strength $E_{0}$
from $0$ to $10^{9}$ V/m. The molecules are assumed to be identical and have transition energy $\hbar\omega_{eg}^{\left(n\right)}=2.6$
eV and transition dipole moments $d_{gf}^{\left(n\right)}=16$
D and $d_{ge}^{\left(n\right)}=14.4$ D. Here, we choose the transition
energy $\hbar\omega_{fg}^{\left(n\right)}=2.7$ eV to avoid
resonant energy transfer to higher multipole plasmons.
The decay rate $\hbar k_{f\to e}$ is varied from $0$ to $100$ meV
while the other rates are zero. When the parameters are varied in the main text,
we state their values in the figure legend.

The amplitude of the driving field $E_{0}$ is related to  power
density $P$ by the relation $P=E^{2}/Z_{0}$ where
$Z_{0}\approx377$ $\Omega$ is characteristic impedance of vacuum.
When $E_{0}$ increases from $10^{6}$, $10^{7}$, $10^{8}$ to $10^{9}$
V/m, $P$ increases from $264$ $\mathrm{kW/cm^{2}}$, $26.4$ $\mathrm{MW/cm^{2}}$,
$2.64\times10^{3}$ $\mathrm{MW/cm^{2}}$ to $2.65\times10^{5}$ $\mathrm{MW/cm^{2}}$.
These values are consistent with the values used in the experiments [20-22,24,28,35,41,45].

\begin{table}
\caption{\label{tab:parameters}Used parameters (for explanation see text)}
\centering{}%
\begin{tabular}{cc|cc}
\hline 
$\hbar\omega_{\mathrm{pl}}$  & $2.6$ eV  & $\hbar\omega_{eg}^{\left(n\right)}$  & $2.6$ eV\tabularnewline
$\hbar\gamma_{\mathrm{pl}}$  & $100$ meV  & $\hbar\omega_{fg}^{\left(n\right)}$  & $2.7$ eV\tabularnewline
$d_{\mathrm{pl}}$  & $2925$ D  & $d_{gf}^{\left(n\right)}$  & $16$ D\tabularnewline
$\Delta x_{\mathrm{mol-MNP}}$  & $2.5$ nm  & $d_{ge}^{\left(n\right)}$  & $14.4$ D\tabularnewline
$E_{0}$  & $0...10^{9}$ ($1.2\times10^{8}$) V/m  & $\hbar k_{f\to e}^{\left(n\right)}$  & $0...100$ ($100$) meV \tabularnewline
$\hbar\omega_{0}$  & $2.7$ eV  & others  & $0$ meV\tabularnewline
\hline 
\end{tabular}
\end{table}

\section{Collective Reduced Density Matrix,  Population and Emission Spectrum\label{sec:population-emission}}

\begin{figure*}
\centering{}
\begin{align}
 & \frac{\partial}{\partial t}\rho_{{\bf n}}^{\mu\nu}=-i\left(\mu-\nu\right)\omega_{\mathrm{pl}}\rho_{{\bf n}}^{\mu\nu}-\left(\gamma_{\mathrm{pl}}/2\right)\left(\left(\mu+\nu\right)\rho_{\mathbf{n}}^{\mu\nu}-2\sqrt{\left(\mu+1\right)\left(\nu+1\right)}\rho_{\mathbf{n}}^{\mu+1\nu+1}\right)\nonumber \\
 & -i\sum_{a\neq b}\omega_{ba}n_{ab}\rho_{{\bf n}}^{\mu\nu}-\sum_{a\neq b}\left(k_{a\to b}/2\right)\left(\sum_{c}\left(n_{ac}+n_{ca}\right)\rho_{\mathbf{n}}^{\mu\nu}-2n_{aa}\rho_{\left(n_{aa}-1,n_{bb}+1\right)}^{\mu\nu}\right)\nonumber \\
 & +iv_{ge}\sum_{a=g,e,f}[n_{ag}\sqrt{\nu}\rho_{\left(n_{ag}-1,n_{ae}+1\right)}^{\mu\nu-1}-n_{ea}\sqrt{\mu+1}\rho_{\left(n_{ea}-1,n_{ga}+1\right)}^{\mu+1\nu}+n_{ae}\sqrt{\nu\text{+1}}\rho_{\left(n_{ae}-1,n_{ag}+1\right)}^{\mu\nu+1}-\sqrt{\mu}n_{ga}\rho_{\left(n_{ga}-1,n_{ea}+1\right)}^{\mu-1\nu}]\nonumber \\
 & +iv_{gf}\sum_{a=g,e,f}[e^{i\omega_{0}t}\left(n_{af}\rho_{\left(n_{af}-1,n_{ag}+1\right)}^{\mu\nu}-n_{ga}\rho_{\left(n_{ga}-1,n_{fa}+1\right)}^{\mu\nu}\right)+e^{-i\omega_{0}t}\left(n_{ag}\rho_{\left(n_{ag}-1,n_{af}+1\right)}^{\mu\nu}-n_{fa}\rho_{\left(n_{fa}-1,n_{ga}+1\right)}^{\mu\nu}\right)].\label{eq:symmetry-rdm}
\end{align}
\end{figure*}

In the main text, we have introduced the collective reduced density matrix (RDM) $\rho_{\bf n}^{\mu \nu}$ and explained the procedure to derive Eq. (\ref{eq:symmetry-rdm}) for such a matrix (see the next page).  More details can be found in \cite{YZhang2}. In Eq. (\ref{eq:symmetry-rdm}) $\omega_{ba}=\left(E_{b_n}-E_{a_n}\right)/\hbar$, $v_{ge}=v_{ge}^{\left(n\right)}$
and $v_{gf}=v_{gf}^{\left(n\right)}$. To abbreviate the notion, we
indicate only the numbers that change, for example, $\left(n_{gg}-1,n_{ee}+1\right)$
represents $\bf n$ (${\bf n}={\left\{ n_{cf}\right\} }$) except $n_{gg}$
is reduced by one and $n_{ee}$ is increased by one. 

In the following, we explain how to calculate the population and the
emission spectrum from the collective RDM.
 The population of the system states $\left|\alpha\mu\right\rangle $  can calculated by $P_{\left(n_{gg},n_{ee},n_{ff}\right)}^{\mu}\equiv\rho_{\left(n_{gg},0,0,0,n_{ee},0,0,0,n_{ff}\right)}^{\mu\mu}$
with $n_{aa}$ being the number of molecules on the states $\left|a_{n}\right\rangle $.
The population of the moleculer product states  $\left|\alpha\right\rangle $ is $P_{\left(n_{gg},n_{ee},n_{ff}\right)}=\sum_{\mu}P_{\left(n_{gg},n_{ee},n_{ff}\right)}^{\mu}$.
The state population of individual molecules $P_{a}=P_{a_n}$ can be calculated with
\begin{align} 
P_{g} & =\sum_{n_{gg}=1}^{N_{\text{m}}}\sum_{n_{ee}=0}^{N_{\text{m}}-n_{gg}}C_{N_{\text{m}}-1}^{n_{gg}-1}C_{N_{\text{m}}-n_{gg}}^{n_{ee}}P_{\left(n_{gg},n_{ee},N_{\text{m}}-n_{gg}-n_{ee}\right)},\label{eq:pg-collective}\\
P_{e} & =\sum_{n_{ee}=1}^{N_{\text{m}}}\sum_{n_{gg}=0}^{N_{\text{m}}-n_{ee}}C_{N_{\text{m}}-1}^{n_{ee}-1}C_{N_{\text{m}}-n_{ee}}^{n_{gg}}P_{\left(n_{gg},n_{ee},N_{\text{m}}-n_{gg}-n_{ee}\right)},\label{eq:pe-collective}\\
P_{f} & =\sum_{n_{ff}=1}^{N_{\text{m}}}\sum_{n_{ee}=0}^{N_{\text{m}}-n_{ff}}C_{N_{\text{m}}-1}^{n_{ff}-1}C_{N_{\text{m}}-n_{ff}}^{n_{ee}}P_{\left(N_{\text{m}}-n_{ee}-n_{ff},n_{ee},n_{ff}\right)},\label{eq:pf-collective}
\end{align}
where $C_{n}^{m}=n!/\left[m!\left(n-m\right)!\right]$ is a combinational coefficient. 
The population of the plasmon states $\left|\mu\right\rangle $ can be calculated from 
\begin{equation}
P_{\mu}=\sum_{n_{gg}=0}^{N_{\text{m}}}\sum_{n_{ee}=0}^{N_{\text{m}}-n_{gg}}C_{N_{\text{m}}}^{n_{gg}}C_{N_{\text{m}}-n_{gg}}^{n_{ee}}P_{\left(n_{gg},n_{ee},N_{\text{m}}-n_{gg}-n_{ee}\right)}^{\mu}.\label{eq:pmu-collective}
\end{equation}
  The average plasmon number $A_{\mathrm{c}}=\sum_{\mu}\mu P_{\mu}$ and
the second order correlation function at steady-state $g^{2}\left(0\right)=\mathrm{tr}_{\textrm{s}}\left\{ C^{+}C^{+}CC\hat{\rho}_\text{ss}\right\} /A_{\mathrm{c}}^{2}=\sum_{\mu}\mu\left(\mu-1\right)P_{\mu}/A_{\mathrm{c}}^{2}$
can be directly calculated from $P_{\mu}$. Here, $\hat{\rho}_\text{ss}$
is the  steady state system RDO.

The steady-state emission spectrum $F\left(\omega\right)\propto\mathrm{Re}\int_{0}^{\infty}d\tau e^{-i\omega\tau}\left\langle C^{+}\left(\tau\right)C\left(0\right)\right\rangle $
is determined by the \emph{Fourier} transformation of the correlation
function $\left\langle C^{+}\left(\tau\right)C\left(0\right)\right\rangle $.
According to the quantum regression theory [62], $\left\langle C^{+}\left(\tau\right)C\left(0\right)\right\rangle \equiv\mathrm{tr}_{\mathrm{s}}\left\{ C^{+}\hat{\sigma}\left(\tau\right)\right\} $
can be calculated with the operator $\hat{\sigma}$ satisfying the same equation as $\hat{\rho}$ with however the initial condition $C\left(0\right)=C\hat{\rho}_\text{ss}$. For the identical molecules the steady-state emission spectrum becomes
\begin{align}
F\left(\omega\right) & \propto\textrm{Re}\int_{0}^{\infty} d\tau e^{-i\omega\tau}\sum_{\mu}\sqrt{\mu} \sum_{n_{ff}=0}^{N_{\text{m}}}  \sum_{n_{ee}=0}^{N_{\text{m}}-n_{ff}}C_{N_{\text{e}}}^{n_{ff}} C_{N_{\text{e}}-n_{ff}}^{n_{ee}}\nonumber \\
\times & \sigma_{\left(N_{\text{e}}-n_{ff}-n_{ee},0,0,0,n_{ee},0,0,0,n_{ff}\right)}^{\mu-1\mu}\left(\tau\right).\label{eq:symmetry-emission}
\end{align}
The sigma matrix elements $\sigma_{\mathbf{n}}^{\mu,\nu}\left(\tau\right)$
satisfy the same equations as $\rho_{\mathbf{n}}^{\mu,\nu}$, cf. Eq. \eqref{eq:symmetry-rdm}, with however the initial condition $\sigma_{\mathbf{n}}^{\mu,\nu}\left(\tau=0\right)=\sqrt{\mu+1}\rho_{\mathbf{n},\text{ss}}^{\mu+1,\nu}$ given by the steady-state collective RDM.

\section{Derivation of Approximate Equation for Plasmon Reduced Density Matrix\label{sec:approximate-prdm}}

Because of the computational effort involved in solving the collective
RDM equation, we can only simulate systems with few molecules. 
To solve the RDO equation (\ref{eq:RDO}) in the main text for many molecules,
 we have presented an approximate 
method based on the plasmon RDM $\rho_{\mu\nu}\equiv\text{tr}_{\text{S}}\left\{ \hat{\rho}\left(t\right)\left|\nu\right\rangle \left\langle \mu\right|\right\}$. The equation for this matrix can be easily derived from Eq. (\ref{eq:RDO}) : 
\begin{align}
 & \frac{\partial}{\partial t}\rho_{\mu\nu}=-i\omega_{\mu\nu}\rho_{\mu\nu}-\gamma_{\mathrm{pl}}\left[\left(\mu+\nu\right)/2\right]\rho_{\mu\nu}\nonumber \\
 & +\gamma_{\mathrm{pl}}\sqrt{\left(\mu+1\right)\left(\nu+1\right)}\rho_{\mu+1\nu+1}\nonumber \\
 & -\sum_{n=1}^{N_{\mathrm{m}}}iv_{ge}^{\left(n\right)}\left(\sqrt{\mu}\rho_{e\mu-1,g\nu}^{\left(n\right)}-\sqrt{\nu}\rho_{g\mu,e\nu-1}^{\left(n\right)}\right)\nonumber \\
 & +\sum_{n=1}^{N_{\mathrm{m}}}iv_{ge}^{\left(n\right)}\left(\sqrt{\nu+1}\rho_{e\mu,g\nu+1}^{\left(n\right)}-\sqrt{\mu+1}\rho_{g\mu+1,e\nu}^{\left(n\right)}\right).\label{eq:plasmonRDM}
\end{align}
This equation depends on the molecule-plasmon correlations $\rho_{a\mu,b\nu}^{\left(n\right)}$.
The equations for these correlations can be again derived from Eq. (\ref{eq:RDO}) in the main text. For the derivation, we introduce the following
replacement 
\begin{eqnarray}
 & - & \left[i\omega_{\mu\nu}-\gamma_{\mathrm{pl}}\left(\mu+\nu\right)/2\right]\rho_{a\mu,b\nu}^{\left(n\right)}\nonumber \\
 & + & \gamma_{\mathrm{pl}}\sqrt{\left(\mu+1\right)\left(\nu+1\right)}\rho_{a\mu+1,b\nu+1}^{\left(n\right)}\to-i\tilde{\omega}_{\mu\nu}\rho_{a\mu,b\nu}^{\left(n\right)}.\label{eq:replacement}
\end{eqnarray}
This removes the dependence of higher excited states of plasmon due
to the plasmon damping without underestimating its influence. This
has been successfully applied in our previous works on lasers with
two-level [60] and three-level molecules [61],
because the differences introduced by the replacement are very small if the higher number states are involved. Here, the complex
transition frequency is defined as 
\begin{equation}
\tilde{\omega}_{\mu\nu}=\omega_{\mu\nu}-i\gamma_{\mathrm{pl}}\left[\left(\mu+\nu\right)/2-\sqrt{\mu\nu}\right].
\end{equation}
Due to the coupling with the driving field, we can introduce the following slowly varying correlations $\tilde{\rho}_{f\mu,g\nu}^{\left(n\right)}\equiv e^{i\omega_{0}t}\rho_{f\mu,g\nu}^{\left(n\right)}$,
$\tilde{\rho}_{g\mu,f\nu}^{\left(n\right)}\equiv e^{-i\omega_{0}t}\rho_{g\mu,f\nu}^{\left(n\right)}$,
$\tilde{\rho}_{f\mu,e\nu-1}^{\left(n\right)}=e^{i\omega_{0}t}\rho_{f\mu,e\nu-1}^{\left(n\right)}$,
$\tilde{\rho}_{e\mu-1,f\nu}^{\left(n\right)}\equiv e^{-i\omega_{0}t}\rho_{e\mu-1,f\nu}^{\left(n\right)}$.
Finally, we get the equations for the correlations. 

First, we present the equations for the population-like correlations
$\rho_{a\mu,a\nu}^{\left(n\right)}$: 
\begin{align}
 & \frac{\partial}{\partial t}\rho_{g\mu,g\nu}^{\left(n\right)}=-i\tilde{\omega}_{\mu\nu}\rho_{g\mu,g\nu}^{\left(n\right)}-\left(k_{g\to f}^{\left(n\right)}+k_{e\to f}^{\left(n\right)}\right)\rho_{g\mu,g\nu}^{\left(n\right)}\nonumber \\
 & +k_{f\to g}^{\left(n\right)}\rho_{f\mu,f\nu}^{\left(n\right)}+k_{e\to g}^{\left(n\right)}\rho_{e\mu,e\nu}^{\left(n\right)}\nonumber \\
 & +iv_{ge}^{\left(n\right)}\left(\sqrt{\nu}\rho_{g\mu,e\nu-1}^{\left(n\right)}-\sqrt{\mu}\rho_{e\mu-1,g\nu}^{\left(n\right)}\right)\nonumber \\
 & -iv_{gf}^{\left(n\right)}\left(\tilde{\rho}_{f\mu,g\nu}^{\left(n\right)}-\tilde{\rho}_{g\mu,f\nu}^{\left(n\right)}\right),\label{eq:rhogg}
\end{align}
\begin{align}
 & \frac{\partial}{\partial t}\rho_{e\mu-1,e\nu-1}^{\left(n\right)}=-i\tilde{\omega}_{\mu-1\nu-1}\rho_{e\mu-1,e\nu-1}^{\left(n\right)}+k_{g\to e}^{\left(n\right)}\rho_{g\mu-1,g\nu-1}^{\left(n\right)}\nonumber \\
 & +k_{f\to e}^{\left(n\right)}\rho_{f\mu-1,f\nu-1}^{\left(n\right)}-\left(k_{e\to g}^{\left(n\right)}+k_{e\to f}^{\left(n\right)}\right)\rho_{e\mu-1,e\nu-1}^{\left(n\right)}\nonumber \\
 & +iv_{ge}^{\left(n\right)}\left(\sqrt{\nu}\rho_{e\mu-1,g\nu}^{\left(n\right)}-\sqrt{\mu}\rho_{g\mu,e\nu-1}^{\left(n\right)}\right),\label{eq:rhoee}
\end{align}
\begin{align}
 & \frac{\partial}{\partial t}\rho_{f\mu,f\nu}^{\left(n\right)}=-i\tilde{\omega}_{\mu\nu}\rho_{f\mu,f\nu}^{\left(n\right)}-\left(k_{f\to g}^{\left(n\right)}+k_{f\to e}^{\left(n\right)}\right)\rho_{f\mu,f\nu}^{\left(n\right)}\nonumber \\
 & +k_{g\to f}^{\left(n\right)}\rho_{g\mu,g\nu}^{\left(n\right)}+k_{e\to f}^{\left(n\right)}\rho_{e\mu,e\nu}^{\left(n\right)}+iv_{gf}^{\left(n\right)}\left(\tilde{\rho}_{f\mu,g\nu}^{\left(n\right)}-\tilde{\rho}_{g\mu,f\nu}^{\left(n\right)}\right).\label{eq:rhoff}
\end{align}
Obviously, these correlations couple with the coherence-like correlations
$\rho_{a\mu,b\nu}^{\left(n\right)}$ with $a\neq b$. Their equations are: 
\begin{align}
 & \frac{\partial}{\partial t}\rho_{g\mu,e\nu-1}^{\left(n\right)}=i\left(\tilde{\omega}_{eg}^{\left(n\right)*}-\tilde{\omega}_{\mu\nu-1}\right)\rho_{g\mu,e\nu-1}^{\left(n\right)}\nonumber \\
 & +iv_{ge}^{\left(n\right)}\left(\sqrt{\nu}\rho_{g\mu,g\nu}^{\left(n\right)}-\sqrt{\mu}\rho_{e\mu-1,e\nu-1}^{\left(n\right)}\right)-iv_{gf}^{\left(n\right)}\tilde{\rho}_{f\mu,e\nu-1}^{\left(n\right)},\label{eq:rhoge}
\end{align}
\begin{align}
 & \frac{\partial}{\partial t}\rho_{e\mu-1,g\nu}^{\left(n\right)}=-i\left(\tilde{\omega}_{eg}^{\left(n\right)}+\tilde{\omega}_{\mu-1\nu}\right)\rho_{e\mu-1,g\nu}^{\left(n\right)}\nonumber \\
 & +iv_{ge}^{\left(n\right)}\left(\sqrt{\nu}\rho_{e\mu-1,e\nu-1}^{\left(n\right)}-\sqrt{\mu}\rho_{g\mu,g\nu}^{\left(n\right)}\right)+iv_{gf}^{\left(n\right)}\rho_{e\mu-1,f\nu}^{\left(n\right)},\label{eq:rhoeg}
\end{align}

\begin{align}
 & \frac{\partial}{\partial t}\tilde{\rho}_{e\mu-1,f\nu}^{\left(n\right)}=-i\left(\tilde{\omega}_{ef}^{\left(n\right)}+\tilde{\omega}_{\mu-1\nu}\right)\tilde{\rho}_{e\mu-1,f\nu}^{\left(n\right)}\nonumber \\
 & +iv_{gf}^{\left(n\right)}\rho_{e\mu-1,g\nu}^{\left(n\right)}-iv_{ge}^{\left(n\right)}\sqrt{\mu}\tilde{\rho}_{g\mu,f\nu}^{\left(n\right)},\label{eq:rhoef}
\end{align}
\begin{align}
 & \frac{\partial}{\partial t}\tilde{\rho}_{f\mu,e\nu-1}^{\left(n\right)}=i\left(\tilde{\omega}_{ef}^{\left(n\right)*}-\tilde{\omega}_{\mu\nu-1}\right)\tilde{\rho}_{f\mu,e\nu-1}^{\left(n\right)}\nonumber \\
 & -iv_{gf}^{\left(n\right)}\rho_{g\mu,e\nu-1}^{\left(n\right)}+iv_{ge}^{\left(n\right)}\sqrt{\nu}\tilde{\rho}_{f\mu,g\nu}^{\left(n\right)},\label{eq:rhofe}
\end{align}
\begin{align}
 & \frac{\partial}{\partial t}\tilde{\rho}_{g\mu,f\nu}^{\left(n\right)}=-i\left(\tilde{\omega}_{gf}^{\left(n\right)}+\tilde{\omega}_{\mu\nu}\right)\tilde{\rho}_{g\mu,f\nu}^{\left(n\right)}\nonumber \\
 & -iv_{ge}^{\left(n\right)}\sqrt{\mu}\tilde{\rho}_{e\mu-1,f\nu}^{\left(n\right)}+iv_{gf}^{\left(n\right)}\left(\rho_{g\mu,g\nu}^{\left(n\right)}-\rho_{f\mu,f\nu}^{\left(n\right)}\right),\label{eq:rhogf}
\end{align}
\begin{align}
 & \frac{\partial}{\partial t}\tilde{\rho}_{f\mu,g\nu}^{\left(n\right)}=i\left(\tilde{\omega}_{gf}^{\left(n\right)*}-\tilde{\omega}_{\mu\nu}\right)\tilde{\rho}_{f\mu,g\nu}^{\left(n\right)}\nonumber \\
 & +iv_{ge}^{\left(n\right)}\sqrt{\nu}\tilde{\rho}_{f\mu,e\nu-1}^{\left(n\right)}-iv_{gf}^{\left(n\right)}\left(\rho_{g\mu,g\nu}^{\left(n\right)}-\rho_{f\mu,f\nu}^{\left(n\right)}\right).\label{eq:rhofg}
\end{align}
In the above expressions, we have also introduced complex transition
frequencies $\tilde{\omega}_{eg}^{\left(n\right)}=\omega_{eg}^{\left(n\right)}-i\gamma_{eg}^{\left(n\right)}$
with the dephasing rate $\gamma_{eg}^{\left(n\right)}\equiv\left(k_{e\to g}^{\left(n\right)}+k_{e\to f}^{\left(n\right)}+k_{g\to e}^{\left(n\right)}+k_{g\to f}^{\left(n\right)}\right)/2$,
$\tilde{\omega}_{ef}^{\left(n\right)}=\omega_{ef}^{\left(n\right)}+\omega_{0}-i\gamma_{ef}^{\left(n\right)}$
with $\gamma_{ef}^{\left(n\right)}\equiv\left(k_{e\to g}^{\left(n\right)}+k_{e\to f}^{\left(n\right)}+k_{f\to g}^{\left(n\right)}+k_{f\to e}^{\left(n\right)}\right)/2$
and $\tilde{\omega}_{gf}^{\left(n\right)}=\omega_{gf}^{\left(n\right)}+\omega_{0}-i\gamma_{gf}^{\left(n\right)}$
with $\gamma_{gf}^{\left(n\right)}\equiv\left(k_{g\to e}^{\left(n\right)}+k_{g\to f}^{\left(n\right)}+k_{f\to e}^{\left(n\right)}+k_{f\to g}^{\left(n\right)}\right)/2$.
If relevant, pure dephasing rate of the molecules can be included in the above dephasing rates.

Following the procedure stated in the main text, we can get closed equations for the plasmon
RDM. In practice, we proceed and first express the steadystate coherence-like
correlations $\rho_{a\mu,b\nu}^{\left(n\right)}$ ($a\neq b$) with
the population-like correlations $\rho_{a\mu,a\nu}^{\left(n\right)}$. Second, we insert those expressions in the equations for
the latter correlations and solve the closed equations analytically.
Third, we utilize the expressions achieved to get analytical solutions for the correlations, which  
we insert to Eq. \eqref{eq:plasmonRDM} to get
the equations for $\rho_{\mu \nu}$. 

By setting the time-derivatives to zero in Eqs. \eqref{eq:rhoef}
and \eqref{eq:rhofe}, we have 
\begin{equation}
\tilde{\rho}_{e\mu-1,f\nu}^{\left(n\right)}=\frac{v_{gf}^{\left(n\right)}\rho_{e\mu-1,g\nu}^{\left(n\right)}-v_{ge}^{\left(n\right)}\sqrt{\mu}\tilde{\rho}_{g\mu,f\nu}^{\left(n\right)}}{\tilde{\omega}_{ef}^{\left(n\right)}+\tilde{\omega}_{\mu-1\nu}},\label{eq:rhofe-1-1}
\end{equation}
\begin{equation}
\tilde{\rho}_{f\mu,e\nu-1}^{\left(n\right)}=\frac{v_{gf}^{\left(n\right)}\rho_{g\mu,e\nu-1}^{\left(n\right)}-v_{ge}^{\left(n\right)}\sqrt{\nu}\tilde{\rho}_{f\mu,g\nu}^{\left(n\right)}}{\tilde{\omega}_{ef}^{\left(n\right)*}-\tilde{\omega}_{\mu\nu-1}}.\label{eq:rhoef-1-1}
\end{equation}
From Eqs. \eqref{eq:rhogf} and \eqref{eq:rhofg} we get 
\begin{equation}
\tilde{\rho}_{g\mu,f\nu}^{\left(n\right)}=\frac{v_{gf}^{\left(n\right)}\left(\rho_{g\mu,g\nu}^{\left(n\right)}-\rho_{f\mu,f\nu}^{\left(n\right)}\right)-v_{ge}^{\left(n\right)}\sqrt{\mu}\tilde{\rho}_{e\mu-1,f\nu}^{\left(n\right)}}{\tilde{\omega}_{gf}^{\left(n\right)}+\tilde{\omega}_{\mu\nu}},\label{eq:rhogf-1-1}
\end{equation}
\begin{equation}
\tilde{\rho}_{f\mu,g\nu}^{\left(n\right)}=\frac{v_{gf}^{\left(n\right)}\left(\rho_{g\mu,g\nu}^{\left(n\right)}-\rho_{f\mu,f\nu}^{\left(n\right)}\right)-v_{ge}^{\left(n\right)}\sqrt{\nu}\tilde{\rho}_{f\mu,e\nu-1}^{\left(n\right)}}{\tilde{\omega}_{gf}^{\left(n\right)*}-\tilde{\omega}_{\mu\nu}}.\label{eq:rhofg-1-1}
\end{equation}
Inserting Eqs. \eqref{eq:rhogf-1-1} and \eqref{eq:rhofg-1-1} in 
Eqs.  \eqref{eq:rhofe-1-1}  and \eqref{eq:rhoef-1-1}, we have 
\begin{align}
 & \tilde{\rho}_{e\mu-1,f\nu}^{\left(n\right)}=\Xi_{\mu\nu}^{\left(n\right)}v_{gf}^{\left(n\right)}[\left(\tilde{\omega}_{gf}^{\left(n\right)}+\tilde{\omega}_{\mu\nu}\right)\rho_{e\mu-1,g\nu}^{\left(n\right)}\nonumber \\
 & -v_{ge}^{\left(n\right)}\sqrt{\mu}\left(\rho_{g\mu,g\nu}^{\left(n\right)}-\rho_{f\mu,f\nu}^{\left(n\right)}\right)],\label{eq:rhoef-1-1-1}
\end{align}
\begin{align}
 & \tilde{\rho}_{f\mu,e\nu-1}^{\left(n\right)}=\tilde{\Xi}_{\mu\nu}^{\left(n\right)}v_{gf}^{\left(n\right)}[\left(\tilde{\omega}_{gf}^{\left(n\right)*}-\tilde{\omega}_{\mu\nu}\right)\rho_{g\mu,e\nu-1}^{\left(n\right)}\nonumber \\
 & -v_{ge}^{\left(n\right)}\sqrt{\nu}\left(\rho_{g\mu,g\nu}^{\left(n\right)}-\rho_{f\mu,f\nu}^{\left(n\right)}\right)].\label{eq:rhofe-1-1-1}
\end{align}
Here, we have introduced the abbreviations 
\begin{align}
1/\Xi_{\mu\nu}^{\left(n\right)} & =\left(\tilde{\omega}_{ef}^{\left(n\right)}+\tilde{\omega}_{\mu-1\nu}\right)\left(\tilde{\omega}_{gf}^{\left(n\right)}+\tilde{\omega}_{\mu\nu}\right)-v_{ge}^{\left(n\right)2}\mu,\\
1/\tilde{\Xi}_{\mu\nu}^{\left(n\right)} & =\left(\tilde{\omega}_{ef}^{\left(n\right)*}-\tilde{\omega}_{\mu\nu-1}\right)\left(\tilde{\omega}_{gf}^{\left(n\right)*}-\tilde{\omega}_{\mu\nu}\right)-v_{ge}^{\left(n\right)2}\nu.
\end{align}
Inserting Eqs. \eqref{eq:rhofe-1-1} and \eqref{eq:rhoef-1-1} in
Eqs. \eqref{eq:rhogf-1-1} and \eqref{eq:rhofg-1-1} , we have 
\begin{align}
 & \tilde{\rho}_{g\mu,f\nu}^{\left(n\right)}=-\Xi_{\mu\nu}^{\left(n\right)}v_{gf}^{\left(n\right)}[v_{ge}^{\left(n\right)}\sqrt{\mu}\rho_{e\mu-1,g\nu}^{\left(n\right)}\nonumber \\
 & -\left(\tilde{\omega}_{ef}^{\left(n\right)}+\tilde{\omega}_{\mu-1\nu}\right)\left(\rho_{g\mu,g\nu}^{\left(n\right)}-\rho_{f\mu,f\nu}^{\left(n\right)}\right)],\label{eq:rhogf-1}
\end{align}
\begin{align}
 & \tilde{\rho}_{f\mu,g\nu}^{\left(n\right)}=-\tilde{\Xi}_{\mu\nu}^{\left(n\right)}v_{gf}^{\left(n\right)}[v_{ge}^{\left(n\right)}\sqrt{\nu}\rho_{g\mu,e\nu-1}^{\left(n\right)}\nonumber \\
 & -\left(\tilde{\omega}_{ef}^{\left(n\right)*}-\tilde{\omega}_{\mu\nu-1}\right)\left(\rho_{g\mu,g\nu}^{\left(n\right)}-\rho_{f\mu,f\nu}^{\left(n\right)}\right)].\label{eq:rhofg-1}
\end{align}

We now insert Eqs.\eqref{eq:rhogf-1} and \eqref{eq:rhofg-1}
into Eqs. \eqref{eq:rhogg} and \eqref{eq:rhoff} and obtain 
\begin{align}
 & i\tilde{\omega}_{\mu\nu}\rho_{g\mu,g\nu}^{\left(n\right)}=k_{f\to g}^{\left(n\right)}\rho_{f\mu,f\nu}^{\left(n\right)}+k_{e\to g}^{\left(n\right)}\rho_{e\mu,e\nu}^{\left(n\right)}+iv_{gf}^{\left(n\right)2}\left(\rho_{g\mu,g\nu}^{\left(n\right)}-\rho_{f\mu,f\nu}^{\left(n\right)}\right)\nonumber \\
 & \times\left[\Xi_{\mu\nu}^{\left(n\right)}\left(\tilde{\omega}_{ef}^{\left(n\right)}+\tilde{\omega}_{\mu-1\nu}\right)-\tilde{\Xi}_{\mu\nu}^{\left(n\right)}\left(\tilde{\omega}_{ef}^{\left(n\right)*}-\tilde{\omega}_{\mu\nu-1}\right)\right]\nonumber \\
 & +iv_{ge}^{\left(n\right)}\left(1+v_{gf}^{\left(n\right)2}\tilde{\Xi}_{\mu\nu}^{\left(n\right)}\right)\sqrt{\nu}\rho_{g\mu,e\nu-1}^{\left(n\right)}\nonumber \\
 & -iv_{ge}^{\left(n\right)}\left(1+v_{gf}^{\left(n\right)2}\Xi_{\mu\nu}^{\left(n\right)}\right)\sqrt{\mu}\rho_{e\mu-1,g\nu}^{\left(n\right)},\label{eq:rhogg-1-1}
\end{align}
\begin{align}
 & \left[i\tilde{\omega}_{\mu\nu}+k_{f\to g}^{\left(n\right)}+k_{f\to e}^{\left(n\right)}\right]\rho_{f\mu,f\nu}^{\left(n\right)}=k_{g\to f}^{\left(n\right)}\rho_{g\mu,g\nu}^{\left(n\right)}+k_{e\to f}^{\left(n\right)}\rho_{e\mu,e\nu}^{\left(n\right)}\nonumber \\
 & +iv_{gf}^{\left(n\right)2}\left(\tilde{\Xi}_{\mu\nu}^{\left(n\right)}\left(\tilde{\omega}_{ef}^{\left(n\right)*}-\tilde{\omega}_{\mu\nu-1}\right)-\Xi_{\mu\nu}^{\left(n\right)}\left(\tilde{\omega}_{ef}^{\left(n\right)}+\tilde{\omega}_{\mu-1\nu}\right)\right)\nonumber \\
 & \times\left(\rho_{g\mu,g\nu}^{\left(n\right)}-\rho_{f\mu,f\nu}^{\left(n\right)}\right)-iv_{gf}^{\left(n\right)2}\tilde{\Xi}_{\mu\nu}^{\left(n\right)}v_{ge}^{\left(n\right)}\sqrt{\nu}\rho_{g\mu,e\nu-1}^{\left(n\right)}\nonumber \\
 & +iv_{gf}^{\left(n\right)2}\Xi_{\mu\nu}^{\left(n\right)}v_{ge}^{\left(n\right)}\sqrt{\mu}\rho_{e\mu-1,g\nu}^{\left(n\right)}.\label{eq:rhoff-1-1}
\end{align}
In addition, we also write down explicitly Eq. \eqref{eq:rhoee} in
steady state as
\begin{align}
 & \left(i\tilde{\omega}_{\mu-1\nu-1}+k_{e\to g}^{\left(n\right)}+k_{e\to f}^{\left(n\right)}\right)\rho_{e\mu-1,e\nu-1}^{\left(n\right)}\nonumber \\
 & =k_{g\to e}^{\left(n\right)}\rho_{g\mu-1,g\nu-1}^{\left(n\right)}+k_{f\to e}^{\left(n\right)}\rho_{f\mu-1,f\nu-1}^{\left(n\right)}\nonumber \\
 & +iv_{ge}^{\left(n\right)}\left(\sqrt{\nu}\rho_{e\mu-1,g\nu}^{\left(n\right)}-\sqrt{\mu}\rho_{g\mu,e\nu-1}^{\left(n\right)}\right).\label{eq:rhoee-1-1}
\end{align}
From Eqs. \eqref{eq:rhoge} and \eqref{eq:rhoeg}, we have 
\begin{align}
 & \rho_{g\mu,e\nu-1}^{\left(n\right)}=\tilde{\Phi}_{\mu\nu}^{\left(n\right)}\tilde{\Xi}_{\mu\nu}^{\left(n\right)}v_{gf}^{\left(n\right)2}v_{ge}^{\left(n\right)}\sqrt{\nu}\left(\rho_{g\mu,g\nu}^{\left(n\right)}-\rho_{f\mu,f\nu}^{\left(n\right)}\right)\nonumber \\
 & +\tilde{\Phi}_{\mu\nu}^{\left(n\right)}v_{ge}^{\left(n\right)}\left(\sqrt{\nu}\rho_{g\mu,g\nu}^{\left(n\right)}-\sqrt{\mu}\rho_{e\mu-1,e\nu-1}^{\left(n\right)}\right),\label{eq:rhoge-2}
\end{align}
\begin{align}
 & \rho_{e\mu-1,g\nu}^{\left(n\right)}=\Phi_{\mu\nu}^{\left(n\right)}\Xi_{\mu\nu}^{\left(n\right)}v_{gf}^{\left(n\right)2}v_{ge}^{\left(n\right)}\sqrt{\mu}\left(\rho_{g\mu,g\nu}^{\left(n\right)}-\rho_{f\mu,f\nu}^{\left(n\right)}\right)\nonumber \\
 & +\Phi_{\mu\nu}^{\left(n\right)}v_{ge}^{\left(n\right)}\left(\sqrt{\mu}\rho_{g\mu,g\nu}^{\left(n\right)}-\sqrt{\nu}\rho_{e\mu-1,e\nu-1}^{\left(n\right)}\right).\label{eq:rhoeg-2}
\end{align}
Here, we have introduced the abbreviations 
\begin{align}
1/\Phi_{\mu\nu}^{\left(n\right)} & =\Xi_{\mu\nu}^{\left(n\right)}v_{gf}^{\left(n\right)2}\left(\tilde{\omega}_{gf}^{\left(n\right)}+\tilde{\omega}_{\mu\nu}\right)-\left(\tilde{\omega}_{eg}^{\left(n\right)}+\tilde{\omega}_{\mu-1\nu}\right),\\
1/\tilde{\Phi}_{\mu\nu}^{\left(n\right)} & =\tilde{\Xi}_{\mu\nu}^{\left(n\right)}v_{gf}^{\left(n\right)2}\left(\tilde{\omega}_{gf}^{\left(n\right)*}-\tilde{\omega}_{\mu\nu}\right)-\left(\tilde{\omega}_{eg}^{\left(n\right)*}-\tilde{\omega}_{\mu\nu-1}\right).
\end{align}
Inserting Eqs. \eqref{eq:rhoge-2} and \eqref{eq:rhoeg-2} to Eqs. \eqref{eq:rhogg-1-1}, \eqref{eq:rhoff-1-1} and \eqref{eq:rhoee-1-1},
we get 
\begin{align}
 & \left[\begin{array}{ccc}
M_{n;\mu\nu}^{\left(11\right)} & M_{n;\mu\nu}^{\left(12\right)} & M_{n;\mu\nu}^{\left(13\right)}\\
M_{n;\mu\nu}^{\left(21\right)} & M_{n;\mu\nu}^{\left(22\right)} & M_{n;\mu\nu}^{\left(23\right)}\\
M_{n;\mu\nu}^{\left(31\right)} & M_{n;\mu\nu}^{\left(32\right)} & M_{n;\mu\nu}^{\left(33\right)}
\end{array}\right]\left[\begin{array}{c}
\rho_{g\mu,g\nu}^{\left(n\right)}\\
\rho_{e\mu-1,e\nu-1}^{\left(n\right)}\\
\rho_{f\mu,f\nu}^{\left(n\right)}
\end{array}\right]\nonumber \\
 & =\left[\begin{array}{c}
k_{e\to g}^{\left(n\right)}\\
-k_{e\to f}^{\left(n\right)}\\
0
\end{array}\right]\rho_{\mu\nu}+\left[\begin{array}{c}
0\\
0\\
-k_{f\to e}^{\left(n\right)}
\end{array}\right]\rho_{\mu-1\nu-1}.\label{eq:matrix-equation}
\end{align}
Here, we have used $\rho_{e\mu,e\nu}^{\left(n\right)}=\rho_{\mu\nu}-\rho_{g\mu,g\nu}^{\left(n\right)}-\rho_{f\mu,f\nu}^{\left(n\right)}$
to get the first and second line of Eq. \eqref{eq:matrix-equation}.
In order to get the third line of Eq. \eqref{eq:matrix-equation},
we replace $\rho_{f\mu-1,f\nu-1}$ by $\rho_{\mu-1,\nu-1}-\rho_{g\mu-1,g\nu-1}^{\left(n\right)}-\rho_{e\mu-1,e\nu-1}^{\left(n\right)}$
and then upgrade $\rho_{g\mu-1,g\nu-1}^{\left(n\right)}$ by $\rho_{g\mu,g\nu}^{\left(n\right)}$.
The matrix elements $M_{n;\mu\nu}^{\left(ij\right)}$ are defined as follows
\begin{align}
M_{n;\mu\nu}^{\left(11\right)} & =i\tilde{\omega}_{\mu\nu}+k_{e\to g}^{\left(n\right)}+A_{\mu\nu}^{\left(n\right)}+2B_{\mu\nu}^{\left(n\right)}+C_{\mu\nu}^{\left(n\right)},\\
M_{n;\mu\nu}^{\left(12\right)} & =-\left(E_{\mu\nu}^{\left(n\right)}+F_{\mu\nu}^{\left(n\right)}\right),\\
M_{n;\mu\nu}^{\left(13\right)} & =-\left(k_{f\to g}^{\left(n\right)}-k_{e\to g}^{\left(n\right)}+B_{\mu\nu}^{\left(n\right)}+C_{\mu\nu}^{\left(n\right)}\right),\\
M_{n;\mu\nu}^{\left(21\right)} & =k_{g\to f}^{\left(n\right)}-k_{e\to f}^{\left(n\right)}+B_{\mu\nu}^{\left(n\right)}+C_{\mu\nu}^{\left(n\right)},\\
M_{n;\mu\nu}^{\left(22\right)} & =-F_{\mu\nu}^{\left(n\right)},\\
M_{n;\mu\nu}^{\left(23\right)} & =-\left(i\tilde{\omega}_{\mu\nu}+k_{f\to g}^{\left(n\right)}+k_{f\to e}^{\left(n\right)}+k_{e\to f}^{\left(n\right)}+C_{\mu\nu}^{\left(n\right)}\right),\\
M_{n;\mu\nu}^{\left(31\right)} & =k_{g\to e}^{\left(n\right)}-k_{f\to e}^{\left(n\right)}+F_{\mu\nu}^{\left(n\right)}+E_{\mu\nu}^{\left(n\right)},\\
M_{n;\mu\nu}^{\left(32\right)} & =-\left(i\tilde{\omega}_{\mu-1\nu-1}+k_{e\to g}^{\left(n\right)}+k_{e\to f}^{\left(n\right)}+k_{f\to e}^{\left(n\right)}-G_{\mu\nu}^{\left(n\right)}\right),
\end{align}
\begin{align}
M_{n;\mu\nu}^{\left(33\right)} & =-F_{\mu\nu}^{\left(n\right)}.
\end{align}
Here, to simplify the expressions, we have introduced the abbreviations
\begin{align}
 & A_{\mu\nu}^{\left(n\right)}=iv_{ge}^{\left(n\right)2}\left(\Phi_{\mu\nu}^{\left(n\right)}\mu-\tilde{\Phi}_{\mu\nu}^{\left(n\right)}\nu\right),\\
 & B_{\mu\nu}^{\left(n\right)}=iv_{gf}^{\left(n\right)2}v_{ge}^{\left(n\right)2}\left(\Phi_{\mu\nu}^{\left(n\right)}\Xi_{\mu\nu}^{\left(n\right)}\mu-\tilde{\Phi}_{\mu\nu}^{\left(n\right)}\tilde{\Xi}_{\mu\nu}^{\left(n\right)}\nu\right),\\
 & C_{\mu\nu}^{\left(n\right)}=iv_{gf}^{\left(n\right)4}v_{ge}^{\left(n\right)2}\left(\Phi_{\mu\nu}^{\left(n\right)}\Xi_{\mu\nu}^{\left(n\right)2}\mu-\tilde{\Phi}_{\mu\nu}^{\left(n\right)}\tilde{\Xi}_{\mu\nu}^{\left(n\right)2}\nu\right)\nonumber \\
 & -iv_{gf}^{\left(n\right)2}\left[\Xi_{\mu\nu}^{\left(n\right)}\left(\tilde{\omega}_{ef}^{\left(n\right)}+\tilde{\omega}_{\mu-1\nu}\right)-\tilde{\Xi}_{\mu\nu}^{\left(n\right)}\left(\tilde{\omega}_{ef}^{\left(n\right)*}-\tilde{\omega}_{\mu\nu-1}\right)\right],
\end{align}
\begin{align}
 & E_{\mu\nu}^{\left(n\right)}=iv_{ge}^{\left(n\right)2}\sqrt{\mu\nu}\left(\Phi_{\mu\nu}^{\left(n\right)}-\tilde{\Phi}_{\mu\nu}^{\left(n\right)}\right),\\
 & F_{\mu\nu}^{\left(n\right)}=iv_{ge}^{\left(n\right)2}v_{gf}^{\left(n\right)2}\sqrt{\mu\nu}\left(\Phi_{\mu\nu}^{\left(n\right)}\Xi_{\mu\nu}^{\left(n\right)}-\tilde{\Phi}_{\mu\nu}^{\left(n\right)}\tilde{\Xi}_{\mu\nu}^{\left(n\right)}\right),\\
 & G_{\mu\nu}^{\left(n\right)}=iv_{ge}^{\left(n\right)2}\left(\tilde{\Phi}_{\mu\nu}^{\left(n\right)}\mu-\Phi_{\mu\nu}^{\left(n\right)}\nu\right).
\end{align}
In the next step, we calculate the inverse matrix of $M_{n;\mu\nu}^{\left(ij\right)}$
and denote it as $O_{n;\mu\nu}^{\left(ij\right)}$ to solve Eq. \eqref{eq:matrix-equation}:
\begin{align}
 & \rho_{g\mu,g\nu}^{\left(n\right)}=\left(O_{n;\mu\nu}^{\left(11\right)}k_{e\to g}^{\left(n\right)}-O_{n;\mu\nu}^{\left(12\right)}k_{e\to f}^{\left(n\right)}\right)\rho_{\mu\nu}\nonumber \\
 & -O_{n;\mu\nu}^{\left(13\right)}k_{f\to e}^{\left(n\right)}\rho_{\mu-1\nu-1},\label{eq:rhogg-1}
\end{align}
\begin{align}
 & \rho_{e\mu-1,e\nu-1}^{\left(n\right)}=\left(O_{n;\mu\nu}^{\left(21\right)}k_{e\to g}^{\left(n\right)}-O_{n;\mu\nu}^{\left(22\right)}k_{e\to f}^{\left(n\right)}\right)\rho_{\mu\nu}\nonumber \\
 & -O_{n;\mu\nu}^{\left(23\right)}k_{f\to e}^{\left(n\right)}\rho_{\mu-1\nu-1},\label{eq:rhoee-1}
\end{align}
\begin{align}
 & \rho_{f\mu,f\nu}^{\left(n\right)}=\left(O_{n;\mu\nu}^{\left(31\right)}k_{e\to g}^{\left(n\right)}-O_{n;\mu\nu}^{\left(32\right)}k_{e\to f}^{\left(n\right)}\right)\rho_{\mu\nu}\nonumber \\
 & -O_{n;\mu\nu}^{\left(33\right)}k_{f\to e}^{\left(n\right)}\rho_{\mu-1\nu-1}.\label{eq:rhoff-1}
\end{align}

Inserting Eqs. \eqref{eq:rhogg-1}, \eqref{eq:rhoee-1} and \eqref{eq:rhoff-1}
to Eqs. \eqref{eq:rhoge-2} and \eqref{eq:rhoeg-2}, we get 
\begin{align}
 & \rho_{g\mu,e\nu-1}^{\left(n\right)}=\tilde{\Phi}_{\mu\nu}^{\left(n\right)}\tilde{\Xi}_{\mu\nu}^{\left(n\right)}v_{gf}^{\left(n\right)2}v_{ge}^{\left(n\right)}\sqrt{\nu}[\left(O_{n;\mu\nu}^{\left(11\right)}-O_{n;\mu\nu}^{\left(31\right)}\right)k_{e\to g}^{\left(n\right)}\nonumber \\
 & -\left(O_{n;\mu\nu}^{\left(12\right)}-O_{n;\mu\nu}^{\left(32\right)}\right)k_{e\to f}^{\left(n\right)}]\rho_{\mu\nu}\nonumber \\
 & -\tilde{\Phi}_{\mu\nu}^{\left(n\right)}\tilde{\Xi}_{\mu\nu}^{\left(n\right)}v_{gf}^{\left(n\right)2}v_{ge}^{\left(n\right)}\sqrt{\nu}\left(O_{n;\mu\nu}^{\left(13\right)}-O_{n;\mu\nu}^{\left(33\right)}\right)k_{f\to e}^{\left(n\right)}\rho_{\mu-1\nu-1}\nonumber \\
 & +\tilde{\Phi}_{\mu\nu}^{\left(n\right)}v_{ge}^{\left(n\right)}[\left(\sqrt{\nu}O_{n;\mu\nu}^{\left(11\right)}-\sqrt{\mu}O_{n;\mu\nu}^{\left(21\right)}\right)k_{e\to g}^{\left(n\right)}\nonumber \\
 & -\left(\sqrt{\nu}O_{n;\mu\nu}^{\left(12\right)}-\sqrt{\mu}O_{n;\mu\nu}^{\left(22\right)}\right)k_{e\to f}^{\left(n\right)}]\rho_{\mu\nu}\nonumber \\
 & -\tilde{\Phi}_{\mu\nu}^{\left(n\right)}v_{ge}^{\left(n\right)}\left(\sqrt{\nu}O_{n;\mu\nu}^{\left(13\right)}-\sqrt{\mu}O_{n;\mu\nu}^{\left(23\right)}\right)k_{f\to e}^{\left(n\right)}\rho_{\mu-1\nu-1}.\label{eq:rhoge-1}
\end{align}
\begin{align}
 & \rho_{e\mu-1,g\nu}^{\left(n\right)}=\Phi_{\mu\nu}^{\left(n\right)}\Xi_{\mu\nu}^{\left(n\right)}v_{gf}^{\left(n\right)2}v_{ge}^{\left(n\right)}\sqrt{\mu}[\left(O_{n;\mu\nu}^{\left(11\right)}-O_{n;\mu\nu}^{\left(31\right)}\right)k_{e\to g}^{\left(n\right)}\nonumber \\
 & -\left(O_{n;\mu\nu}^{\left(12\right)}-O_{n;\mu\nu}^{\left(32\right)}\right)k_{e\to f}^{\left(n\right)}]\rho_{\mu\nu}\nonumber \\
 & -\Phi_{\mu\nu}^{\left(n\right)}\Xi_{\mu\nu}^{\left(n\right)}v_{gf}^{\left(n\right)2}v_{ge}^{\left(n\right)}\sqrt{\mu}\left(O_{n;\mu\nu}^{\left(13\right)}-O_{n;\mu\nu}^{\left(33\right)}\right)k_{f\to e}^{\left(n\right)}\rho_{\mu-1\nu-1}\nonumber \\
 & +\Phi_{\mu\nu}^{\left(n\right)}v_{ge}^{\left(n\right)}[\left(\sqrt{\mu}O_{n;\mu\nu}^{\left(11\right)}-\sqrt{\nu}O_{n;\mu\nu}^{\left(21\right)}\right)k_{e\to g}^{\left(n\right)}\nonumber \\
 & -\left(\sqrt{\mu}O_{n;\mu\nu}^{\left(12\right)}-\sqrt{\nu}O_{n;\mu\nu}^{\left(22\right)}\right)k_{e\to f}^{\left(n\right)}]\rho_{\mu\nu}\nonumber \\
 & -\Phi_{\mu\nu}^{\left(n\right)}v_{ge}^{\left(n\right)}\left(\sqrt{\mu}O_{n;\mu\nu}^{\left(13\right)}-\sqrt{\nu}O_{n;\mu\nu}^{\left(23\right)}\right)k_{f\to e}^{\left(n\right)}\rho_{\mu-1\nu-1}.\label{eq:rhoeg-1}
\end{align}
With these results, we are now ready to consider the following summations 
\begin{align}
 & iv_{ge}^{\left(n\right)}\left(\sqrt{\mu}\rho_{e\mu-1,g\nu}^{\left(n\right)}-\sqrt{\nu}\rho_{g\mu,e\nu-1}^{\left(n\right)}\right)\nonumber \\
 & =H_{\mu\nu}^{\left(n\right)}\rho_{\mu\nu}-I_{\mu\nu}^{\left(n\right)}\rho_{\mu-1\nu-1},\label{eq:sum-1} \\
 & iv_{ge}^{\left(n\right)}\left(\sqrt{v}\rho_{e\mu-1,g\nu}^{\left(n\right)}-\sqrt{\mu}\rho_{g\mu,e\nu-1}^{\left(n\right)}\right)\nonumber \\
 & =J_{\mu\nu}^{\left(n\right)}\rho_{\mu\nu}-K_{\mu\nu}^{\left(n\right)}\rho_{\mu-1\nu-1}.\label{eq:sum-2}
\end{align}
Here, we have introduced the following abbreviations 
\begin{align}
 & H_{\mu\nu}^{\left(n\right)}=\left(\left(A_{\mu\nu}^{\left(n\right)}+B_{\mu\nu}^{\left(n\right)}\right)O_{n;\mu\nu}^{\left(11\right)}-E_{\mu\nu}^{\left(n\right)}O_{n;\mu\nu}^{\left(21\right)}-B_{\mu\nu}^{\left(n\right)}O_{n;\mu\nu}^{\left(31\right)}\right)k_{e\to g}^{\left(n\right)}\nonumber \\
 & -\left(\left(A_{\mu\nu}^{\left(n\right)}+B_{\mu\nu}^{\left(n\right)}\right)O_{n;\mu\nu}^{\left(12\right)}-E_{\mu\nu}^{\left(n\right)}O_{n;\mu\nu}^{\left(22\right)}-B_{\mu\nu}^{\left(n\right)}O_{n;\mu\nu}^{\left(32\right)}\right)k_{e\to f}^{\left(n\right)},
 \end{align}
 \begin{align}
 & I_{\mu\nu}^{\left(n\right)}=\left(\left(A_{\mu\nu}^{\left(n\right)}+B_{\mu\nu}^{\left(n\right)}\right)O_{n;\mu\nu}^{\left(13\right)}-E_{\mu\nu}^{\left(n\right)}O_{n;\mu\nu}^{\left(23\right)}-B_{\mu\nu}^{\left(n\right)}O_{n;\mu\nu}^{\left(33\right)}\right)k_{f\to e}^{\left(n\right)},
\end{align}
\begin{align}
 & J_{\mu\nu}^{\left(n\right)}=\left(\left(E_{\mu\nu}^{\left(n\right)}+F_{\mu\nu}^{\left(n\right)}\right)O_{n;\mu\nu}^{\left(11\right)}+G_{\mu\nu}^{\left(n\right)}O_{n;\mu\nu}^{\left(21\right)}-F_{\mu\nu}^{\left(n\right)}O_{n;\mu\nu}^{\left(31\right)}\right)k_{e\to g}^{\left(n\right)} \\
 & -\left(\left(E_{\mu\nu}^{\left(n\right)}+F_{\mu\nu}^{\left(n\right)}\right)O_{n;\mu\nu}^{\left(12\right)}+G_{\mu\nu}^{\left(n\right)}O_{n;\mu\nu}^{\left(22\right)}-F_{\mu\nu}^{\left(n\right)}O_{n;\mu\nu}^{\left(32\right)}\right)k_{e\to f}^{\left(n\right)}, \nonumber \\
& K_{\mu\nu}^{\left(n\right)}=\left(\left(E_{\mu\nu}^{\left(n\right)}+F_{\mu\nu}^{\left(n\right)}\right)O_{n;\mu\nu}^{\left(13\right)}+G_{\mu\nu}^{\left(n\right)}O_{n;\mu\nu}^{\left(23\right)}-F_{\mu\nu}^{\left(n\right)}O_{n;\mu\nu}^{\left(33\right)}\right)k_{f\to e}^{\left(n\right)}.
\end{align}
Inserting Eqs. \eqref{eq:sum-1} and \eqref{eq:sum-2} to Eq. \eqref{eq:plasmonRDM},
we get the equations for the plasmon RDM only: 
\begin{align}
 & \frac{\partial}{\partial t}\rho_{\mu\nu}=-i\omega_{\mu\nu}\rho_{\mu\nu}-\gamma_{\mathrm{pl}}\left[\left(\mu+\nu\right)/2\right]\rho_{\mu\nu}\nonumber \\
 & +\gamma_{\mathrm{pl}}\sqrt{\left(\mu+1\right)\left(\nu+1\right)}\rho_{\mu+1\nu+1}\nonumber \\
 & -\sum_{n=1}^{N_{\mathrm{m}}}\left(H_{\mu\nu}^{\left(n\right)}+K_{\mu+1\nu+1}^{\left(n\right)}\right)\rho_{\mu\nu}+\sum_{n=1}^{N_{\mathrm{m}}}I_{\mu\nu}^{\left(n\right)}\rho_{\mu-1\nu-1}\nonumber \\
 & +\sum_{n=1}^{N_{\mathrm{m}}}J_{\mu+1\nu+1}^{\left(n\right)}\rho_{\mu+1\nu+1}.\label{eq:plasmonRDM-final}
\end{align}

The diagonal element of the plasmon RDM can be interpreted as the
plasmon state population $P_{\mu}=\rho_{\mu\mu}$. The equations for
the population can be easily achieved from Eq. \eqref{eq:plasmonRDM-final}
and are given as Eq.  \eqref{eq:plasmon-state-population} in the main text. There, we have introduced the molecule-induced plasmon damping rate
\begin{equation}
k_{\mu}=\sum_{n=1}^{N_{\mathrm{m}}}H_{\mu\mu}^{\left(n\right)}=\sum_{n=1}^{N_{\mathrm{m}}}J_{\mu\mu}^{\left(n\right)},\label{eq:moleculePD}
\end{equation}
and the molecule-induced plasmon pumping rate 
\begin{equation}
p_{\mu}=\sum_{n=1}^{N_{\mathrm{m}}}I_{\mu\mu}^{\left(n\right)}=\sum_{n=1}^{N_{\mathrm{m}}}K_{\mu\mu}^{\left(n\right)}.\label{eq:moleculePP}
\end{equation}

%%%%%%%%%%%%%%%%%%%%%%%%%%%%%%%%%%%
%% 								bibliography 
%%%%%%%%%%%%%%%%%%%%%%%%%%%%%%%%%%%

\end{document}